\documentclass[a4paper, 12pt]{article}

\usepackage{graphicx}  
\usepackage{dcolumn}   
\usepackage{bm}        
\usepackage{amssymb}   

\usepackage[utf8]{inputenc}
\usepackage{slashed}
\usepackage{epsfig}
\usepackage{amsmath,amsfonts,mathtools}
\usepackage{color}
\newcommand{\myincludegraphics}[3]{\hspace{-7pt}
\mkern10mu \raisebox{- #1 pt}{\scalebox{#2}{\includegraphics{#3}}}}

\def\beq{\begin{eqnarray}}    
\def\eeq{\end{eqnarray}}       
\def\nn{\nonumber}               
\def\n{\label}               

\def\ln{\,\mbox{ln}}                  
\def\tr{\,\mbox{tr}}                  
\def\Tr{\,\mbox{Tr}}                  

\def\D{\Delta_g}       

\def\al{\alpha}
\def\be{\beta}

\def\ga{\gamma}
\def\de{\delta}
\def\ep{\epsilon}

\def\na{\nabla}
\def\pa{\partial}

\def\ph{\varphi}

\def\Ga{\Gamma}
\def\De{\Delta}
\def\La{\Lambda}

\begin{document}

\begin{center}
{\Large\bf Gravitational form factors and decoupling in $2D$
}
\vskip 5mm

{\bf Tiago G. Ribeiro$^{a}$\footnote{
E-mail address: \ tgribeiro@fisica.ufjf.br},
\  \   Ilya L. Shapiro$^{a,b,c}$\footnote{
E-mail address: \ shapiro@fisica.ufjf.br},
\  \   Omar Zanusso}$^{d}$\footnote{
E-mail address: \
omar.zanusso@uni-jena.de}
\vskip 3mm

$^{a}$ {\sl Departamento de Física, ICE,
Universidade Federal  de Juiz  de  Fora
\\
Juiz de Fora, 36036-100, Minas Gerais, Brazil}
\vskip 2mm

$^{b}$ {\sl Tomsk State Pedagogical University,
Tomsk, 634041, Russia}
\vskip 2mm

$^{c}$ {\sl Tomsk State University,
Tomsk, 634050, Russia}
\vskip 2mm

$^{d}$ {\sl Theoretisch-Physikalisches Institut, Friedrich-Schiller-Universit\"{a}t Jena,
\\
Max-Wien-Platz 1, 07743 Jena, Germany}
\end{center}
\vskip 4mm

\centerline{{\bf Abstract}}
\vskip 2mm

\begin{quotation}
\noindent
We calculate and analyse non-local gravitational form factors induced by
quantum matter fields in curved two-dimensional space.
The calculations are performed for scalars, spinors and massive vectors
by means
of the covariant heat kernel method up to the second order in the curvature
and confirmed using Feynman diagrams.
The analysis of the
ultraviolet (UV) limit reveals a generalized
``running'' form of the Polyakov action for a nonminimal scalar field
and the usual Polyakov action in the conformally invariant cases.
In the infrared (IR) we establish the
gravitational decoupling theorem, which can be seen directly from
the form factors or from the physical beta function for fields
of any spin.
\vskip 2mm

\vskip 2mm

\end{quotation}


\section{Introduction}

The decoupling of quantum massive field at the energies which are
much smaller than the mass of the field is the cornerstone of the
effective approach to quantum field theory. The application of this
idea to gravity is a subject which attracts a special interest (see,
e.g., \cite{Burgess:2003jk} for a recent review). Practical
calculations of decoupling of massive matter fields in four
dimensional ($4D$) spacetime have been done by means of Feynman
diagrams and also within the heat kernel methods
\cite{Gorbar:2002pw,Gorbar:2003yt,Codello:2012kq}.

There is an extensive literature about the use of effective
field theory methods in quantum gravity. Starting from the pioneer
publications \cite{Iwa,Don94}, there were a lot of works (see, e.g.,
\cite{Don-review} for reviews and further references). Along with
the rich theoretical contents, these works pave the way for
interesting applications in astrophysics. An important point in the
application of effective methods to quantum gravity is the ``belief''
in the universality of quantum general relativity as an IR limit of
quantum gravity of any sort, at least for the local versions of the
theory. The calculation which could resolve this issue has been
proposed long ago by one of the present authors \cite{Gauss,Polemic}.
However, a practical realization of this program meets serious
obstacles, especially because of the technical difficulties of making
calculations in higher derivative gravity within a mass-dependent
renormalization scheme. In this situation it is tempting to start
from the simplest model of quantum gravity, and in this respect the
two dimensional ($2D$) quantum gravity is a very useful model to
test general concepts.

In the present work we start the investigation of decoupling of
massive degrees of freedom in $2D$. In order to explore
the mechanisms and special features of two dimensional gravity,
we consider the loops of matter fields and thus extend the results
which were obtained in  $4D$. The nice feature of the gravitational
contributions of matter fields in two dimensions is that the ones of
the massless conformal fields are pretty well known from the
integration of the trace anomaly \cite{DDI76,Duff77}, being
covariant non-local Polyakov action \cite{Polyakov}. The integration
of non-conformal and in particular massive fields gives a new
perspective to these classical results about anomaly \cite{birdav}
and in this sense it complements the recent derivation of the
Polyakov action via the functional renormalization group methods
\cite{A. Codello:2010}, which effectively treat all fields as
massive due to the use of the regulator term.

In what follows we summarize the derivation of non-local form
factors in the vacuum gravitational sector of massive matter fields
in $2D$. Regardless of its relative simplicity compared to the
$4D$ vacuum calculations, we shall observe that the $2D$ results
may be interesting, since they enable one to explore general issues
as the relation between  massless limit and the anomaly-induced
effective action in the UV from one side and the decoupling of
massive degrees of freedom at the IR end of the energy scale.

The paper is organized as follows. In Sec.~\ref{s2} we present a
brief review of the necessary results from the heat kernel methods,
based on the works \cite{bavi87,bavi90}, \cite{Avramidi:2007zz} and
\cite{Codello:2012kq}. Some useful technical information  is
separated into Appendix \ref{appendix-heat-kernel}, and an explanation of how the surface
term can be dealt with using Feynman diagrams into Appendix \ref{appendix-feynman}.
Sec.~\ref{s3} gives general well-known expressions for the vacuum
effective action of a theory which included sets of massive scalars,
fermions and vectors.
Sec.~\ref{s4} describes the derivation of form factors of a
massive scalar for the gravitational terms in $2D$.  Furthermore,
in the two subsections we describe the generalization of the
Polyakov action to the case of a nonminimal massless scalar
and the high-energy (UV) and low-energy (IR) limits in the
physical beta functions derived within the momentum-subtraction
renormalization scheme. Technical discussion of the result for the
non-conformal case can be found in Appendix \ref{appendix-non-analyticity}.
Sec.~\ref{s5} shows the derivation and
discuss the structure of the form factor and beta functions for the
Dirac fermion. In Sec.~\ref{s6} we show the same for the Proca
field and also discuss the discontinuity which takes place in the
UV, in which one finds a discontinuity with the expression for the
massless gauge field. Finally, in Sec.~\ref{s7} we draw our
conclusions and outline perspectives.

\section{Effective action in $2D$: general considerations}
\label{s2}

In $2D$ Riemann and Ricci tensors can be expressed via the
Ricci scalar, as
\beq\label{rrr}
R_{\mu\nu\al\be} =  \frac12\,R\big(g_{\mu\al}g_{\nu\be} -
g_{\nu\al}g_{\mu\be}\big),
\qquad
R_{\mu\nu} =  \frac12\,R\,g_{\mu\nu}.
\eeq
This means, in particular, that the Einstein
tensor is identically zero such that the part of Einstein's classical
equations describing the metric tensor is null. Furthermore, the
power counting in $2D$ shows much less divergences. Both features
make the $2D$ quantum gravity models much easier to explore
compared than their $4D$ counterparts \cite{Ichinose,OdSh-90ies},
and produce results which are compatible with the ones of
conformal field theory \cite{Knizhnik:1988ak,Distler:1988jt}. Specifically, all
these works were devoted to the quantization of a nonlocal Polyakov
action or its metric-scalar equivalents.

In a $2D$ semi-classical approach, in which only matter fields are
quantized, the effective action for the metric may be induced by
quantum fields. In the UV limit such an  effective action can be
calculated through the integration
of the conformal anomaly \cite{Mottola:1995sj}.However, in order to describe the phenomena at
various energy scales, one has to take the masses of quantum
fields into account. Consider a free massive matter field $\Phi$
characterized by the quadratic action
\beq
 S[\Phi]
&=& \frac{1}{2}\int {\rm d}^D x \sqrt{g}\,\, \Phi \left( {\cal O}
+ m^2 \right) \Phi ,
\eeq
in which ${\cal O}$ is a second-order covariant differential operator
also known as a Laplace type operator.

We can use the heat kernel expansion to compute the
one-loop effective action of vacuum by appropriately integrating
over the heat kernel time in the Euclidean signature,
\beq
 \Gamma[g]
 \,=\, -\frac{1}{2} \Tr
 \int_0^\infty \frac{{\rm d}s}{s} \,\,{\rm e}^{-sm^2} \,{\cal H}(s) ,
\label{gamma}
\eeq
in which we introduced the heat kernel solution ${\cal H}(s)$,
which formally is a bi-scalar ${\cal H}(s)={\cal H}(s;x,x')$
and whose explicit form will be shown in Appendix \ref{appendix-heat-kernel} following \cite{Codello:2012kq,bavi90}.
Generically, he functional trace includes taking coincidence limit $x'\to x$
and a covariant integration over $\int {\rm d}^D x \sqrt{g}$.
Additionally, since the parameter $s$ is dual to an energy
the $s$-integration ranges from the UV at $s=0$ to the IR at $s=\infty$. While the
IR limit converges thanks to the presence of the mass
$m^2>0$, the UV might require regularization and could induce a
running in the gravitational couplings for the effective action $\Gamma[g]$ through renormalization.

In fact, along with a finite non-local part, the effective action includes
divergences. In dimensional regularization these
appear through integrals of the form \cite{bavi85}
\beq
 -\,\frac{1}{2} \int_0^\infty\frac{{\rm d}s}{s^{n+1}}
 \,{\rm e}^{-sm^2} \,\,\sim \,\,\frac{1}{2 n} + {\rm finite}\,.
\eeq
From the form of the heat kernel expansion and considering only the
small-$s$ asymptotics we observe that for $D=2-\ep$ the potential
divergences can appear as poles $\,1/\ep\,$ up to the first order in the
curvature expansion when $n=D/2-1$.
If instead $D=4-\ep$ the divergences could appear up to the second order in the
curvature expansion when $n=D/2-2$.

In order to cancel the divergences one has to introduce counterterms
$\De S[g;\mu]$,  then the renormalized one-loop effective action
becomes
\beq
 \Gamma_{\rm ren}[g] \,=\, \Gamma[g] \,+\, \De S[g;\mu]\,,
\eeq
in which $\mu$ is the mass scale introduced for dimensional reasons 
and at which the subtraction of the divergences occurs.
As long as we are concerned with free fields, the counter terms depend only on
the metric and hence only the parameters of the vacuum action
may be running with scale.

In the minimal subtraction ($\overline{\rm MS}$) scheme of
renormalization the counter terms simply remove the poles $1/\ep$
(modulo some local finite part). In this scheme the subtraction is
performed at the arbitrarily high scale $\mu$ and the effects of masses
on the running of effective parameters is lost. If otherwise the
subtraction is performed at a physical energy scale, say $q^2$, then the
effects that masses have in the IR become visible in the running
of the vacuum parameters. The two extreme regimes of interest
are the UV limit $q^2/m^2\gg 1$ and the IR limit $q^2/m^2\ll 1$, in
which the fluctuations freeze-out below a threshold defined by the
mass of the quantum field.

Explicit calculations in $4D$ \cite{Gorbar:2002pw,Gorbar:2003yt} have
shown that the physical running agrees with the expectations based on the
Appelquist-Carazzone theorem \cite{AC}, namely the beta function $\beta$ of
any vacuum parameter displays the two limits
\beq
&&          \beta_{\rm UV} \,=\, \beta_{\overline{\rm MS}}
          \quad\,\,\,\qquad {\rm for\,\,} q^2/m^2\gg 1
\nn
\\
&&          \beta_{\rm IR} \,\propto\, \frac{q^2}{m^2}
          \qquad \qquad {\rm for\,\,} q^2/m^2\ll 1\,.
\n{UVIR}
\eeq
In the next sections we explore the status of the theorem in the $2D$ case.

\section{Vacuum sector of a general theory at one loop}
\label{s3}

It is always interesting to keep in mind a general model of interacting
matter fields. At one loop the vacuum form factor does not depend on the
interactions and is given by the algebraic sum of the contributions
of the fields with spins $\,0,\,1/2,\,1\,$ with different masses. Let
us consider the vacuum form factors in the theory with $n_{\rm s}$
minimally or nonminimally coupled scalar fields, $n_{\rm f}$
minimally coupled Dirac spinors, and $n_{\rm p}$ minimally coupled
Proca fields on a general $2D$ background with metric $g_{\mu\nu}$.
Instead of massive Proca fields we could consider massless vectors, and
later on we shall see the difference between this case and the
massless limit of the Proca model.

For the sake of simplicity, let us assume that all matter fields of
each spin have the same masses. Then the effective action is given
by the expression
\beq
\Ga[g] &=&
 \frac{n_{\rm s}}{2}\, \Tr_{\rm s} \ln \left(- \D + \xi R +m_{\rm s}^2\right)
\,-\, n_{\rm f} \Tr_{\rm f} \ln \left(\slashed D+m_{\rm f}\right)
\nn
\\
&&
+\,\frac{n_{\rm p}}{2}\,\Tr_{\rm v} \ln\left( - \D + m_{\rm v}^2\right).
\eeq
In this formula we denoted $\D=\nabla^2$ independently of the type
of bundle (scalar, spinor or vector), and we assume that the functional traces
must be taken accordingly.
The first term includes the scalar contributions and, for our current purposes,
it does not need further manipulation, let us instead consider the other two terms.

The Dirac operator $\slashed D$ could be squared inside the trace to obtain
\beq
\Tr_{\rm f} \ln  \left(\slashed D+m_{\rm f}\right)
\,=\,
\frac12\, \Tr_{\rm f}\ln \Big(- \D + \frac{R}{4} + m_{\rm f}^2\Big)
\eeq
in which we used an explicit form for the spin connection.
The latter is also used to evaluate the commutator of covariant derivatives over the spin bundle
\beq
[\na_\mu,\na_\nu] \,=\,\Omega_{\mu\nu}
\,=\,-\,\frac{1}{4}\,\ga^\al\ga^\be R_{\al\be\mu\nu}.
\eeq

The trace constituting the contribution of the Proca fields can be rewritten as the
difference between a vector trace and a scalar trace. We refer to
\cite{bavi85} and \cite{BuGui} for two detailed and different derivations of this property
in curved space. Since these derivations do not depend explicitly on the spacetime dimension
we shall not present the details here, but just give the final result
\beq\label{vect}
\frac{1}{2}\, \Tr_{\rm v} \ln \left( -\D+m_{\rm v}^2\right)
 = \frac{1}{2} \Tr  \ln \left( - g_{\mu\nu}\D + R_{\mu\nu}
 +  g_{\mu\nu} m_{\rm v}^2\right)
 - \frac12 \Tr \ln\left(-\D +m_{\rm v}^2\right)\,
\eeq
The second term is equivalent to a minimally coupled scalar field
with $\xi = 0$.

The final result of all the above manipulations is
\beq
\Ga[g]
&=&
\frac{n_{\rm s}}{2} \Tr_{\rm s} \log\left(-\D + \xi R +m_{\rm s}^2\right)
- \frac{n_{\rm f}}{2} \Tr_{\rm f} \log \left(- \D + \frac{R}{4} - m_{\rm f}^2\right)
\nn
\\
&&+
\frac{n_{\rm p}}{2} \Tr_{\rm v} \log\left( -\D+{\rm Ric}+m_{\rm v}^2\right)
- \frac{n_{\rm p}}{2} \Tr_{\rm s} \log\left(-\D +m_{\rm v}^2\right).
\n{total}
\eeq
In the next sections we derive the above contributions
following the methods anticipated in Sec.~\ref{s2} with the
appropriate replacements of the masses of the fields.
We anticipate the notation
$\be_{G}^{\rm\,s}$, $\be_{G}^{\rm\,f}$, $\be_{G}^{\rm\,p}$
and $\be_{G}^{\rm\,g}$ for the contributions to the running of the
inverse Newton constant induced by scalar, fermionic, Proca and
massless vector gauge degrees of freedom, respectively \cite{Gorbar:2002pw}.


\section{Massive scalar field in $2D$ gravity}
\label{s4}

We begin our computation of the induced vacuum effective action
by considering the effects induced by a massive non-minimally coupled scalar field
throughout the entirety this section.
We consider this to be the most essential example and therefore
we use it to flesh out all the details of our computation.
Consider a scalar field $\ph$ with
the classical action
\beq
 S[\ph] &=&
 \frac{1}{2}\int {\rm d}^2 x \sqrt{g} ~
 \ph \left( -\D + \xi R + m^2\right) \ph.
\n{scal}
\eeq
In general this action is not conformally invariant, but it is well-known
that classically the invariance can be recovered
in the simultaneous limit of $m^2 \to 0$ and
$\xi \to 0$.
More generally, in this limit one could evaluate the effective action $\Ga[g]$ by
integrating the conformal anomaly, because no conformal invariant
structures can be expected in $2D$. The result of this procedure
is the Polyakov term which is a non-local action that is quadratic in the Ricci scalar.
Outside the protection of the conformal limit one can instead
expect all powers of Ricci scalar and more
complicated structure of non-localities. However, for the sake of
practical calculations, we will consider only terms up to the
second order in the curvature of the non-local effective action.

On the top of the above considerations we have to introduce the action of vacuum,
which should be local, covariant and sufficient to
renormalize all possible divergences. In $2D$
such an Euclidean action includes only Einstein-Hilbert and cosmological terms,
\beq
 S_{\rm vac}[g_{\mu\nu}] &=&
 \frac{1}{16\pi G_{\rm N}}\int {\rm d}^2 x \sqrt{g}\,
 \big(2\La-R  \big).
\n{vac}
\eeq
Indeed, the above vacuum action is sufficient for all types of matter
fields, so the renormalization of vacuum reduce to the renormalization
of the Newton constant $G_{\rm N}$ and the cosmological constant
$\La$.

\subsection{Derivation of effective action and $\be$-function}
\label{s4.1}

The regularized effective action is defined as modified version
of the general expression \eqref{gamma},
\beq
 \Ga[g]
 &=&
 - \frac{1}{2}
 \left(4\pi\mu^2\right)^{\ep/2} \,\Tr \int \frac{{\rm d} s}{s}\,\,{\rm e}^{-sm^2}
 \, {\cal H}(s)\,.
\eeq
Here $\mu$ is the renormalization parameter, which is used to
preserve the dimension in $D=2-\ep$ dimension.

We now have to evaluate the heat kernel ${\cal H}(s)$
using the methods of Appendix \ref{appendix-heat-kernel}.
Since we are interested in the limit $\ep \to 0$, we use the relations
\eqref{rrr} for the evaluation of the effective action.
In this case the trace of the heat kernel
simplifies considerably and can be written as
\beq
\Tr ~ {\cal H}(s) &=& \frac{1}{(4\pi s)^{D/2}}
\int {\rm d}^2 x \sqrt{g}\, \Bigl\{1 + s \Bigl[G_R(-s\D)
+\xi G_E(-s\D)\Bigr] R
\nonumber
\\
&&+
s^2 R \Bigl[F_R(-s\D)+\frac{1}{2}F_{Ric}(-s\D)
+ \xi F_{RE}(-s\D)
\nonumber
\\
&&
+ \xi^2 F_E(-s\D) \Bigr]R \Bigl\}
+ {\cal O}\left({\cal R}^3\right)\,,
\label{hk}
\eeq
in which the functions
$G_R$, $G_E$, $F_R$, $F_{Ric}$, $F_{RE}$ and $F_E$
are given in Appendix \ref{appendix-heat-kernel}.
Let us stress once more that this formula is essentially
simpler than its $4D$ counterpart thanks to the use of the $2D$ relations
\eqref{rrr} and thanks to the fact that the commutator of covariant
derivatives $\Omega_{\mu\nu}=0$ when acting on scalars. At the
same time the general dimension $D$ appears in the intermediate
formula \eqref{hk} indicating the use of dimensional regularization.

It is convenient to introduce a condensed notation which simplifies both divergent and finite
parts of the effective action. We define
\beq
&&
\frac{1}{\bar{\ep}}
= \frac{2}{\ep} + \ln \Big(\frac{4\pi\mu^2}{m^2}\Big) - \ga,
\nn
\\
&&
z = -\frac{\D}{m^2},
 \quad
 a = \sqrt{\frac{4z}{4+z}},
 \quad
Y = 1-\frac{1}{a} \log\left|{\frac{1+a/2}{1-a/2}}\right|\,,
\eeq
in which $\gamma$ is the Euler-Mascheroni constant.
Taking the integral over the proper time variable $s$, after some manipulations
we obtain the result for the effective action up to the second order in the scalar curvature
\beq\label{gascal}
\Ga_{\rm s}[g] &=&
\frac{1}{4\pi}\int {\rm d}^2 x \sqrt{g}\, \Big\{
\frac{m^2}{2\,\bar{\epsilon}}+\frac{m^2}{2}
+ \frac{1}{2\,\bar{\epsilon}} \left(\xi-\frac{1}{6}\right) R
+\frac{B(z)}{2}R
+  \frac{1}{24} R\, \frac{C(z)}{\D} \, R
 \Big\}\,,
\eeq
in which we defined the functions
\beq
 B(z)
 &=&
 \frac{1}{18}+2\Big(\xi-\frac{1}{4}\Big)Y+\frac{2Y}{3a^2},
 \nonumber
 \\
 C(z)
 &=&
 -\frac{1}{2}-\frac{6 Y}{a^2}+3 (1-4 \xi) Y
 +\frac{3}{8} a^2 (1-4 \xi )^2 (1-Y)\,.
\n{GC}
\eeq

Finally, subtracting the divergence at the scale defined by Euclidean
momentum $q$, we can obtain the beta function for the inverse Newton
constant. The physical beta function can be computed by acting with
the derivative
${\rm d}/ {\rm d}\ln (q/q_0)$ on the coefficient of the Ricci scalar in the finite part
of the effective action. Following the strategy described in
\cite{Gorbar:2002pw,Gorbar:2003yt}, one can perform the computation directly
in coordinate space by trading $q^2 \leftrightarrow -\D$ in the
final expression. In this way we obtain
\beq\label{betazero}
\beta_{G}^{\rm \,s} &=&
\frac{1}{4\pi} z \, B'(z)
\,=\,   \frac{1}{8\pi}\Bigl[
   -\frac{1}{6}
   +(1-2 \xi ) Y
   -\frac{2 Y}{a^2}
   + \frac{1}{8} (1-4 \xi )a^2 (1-Y)
 \Bigr]\,.
\eeq
One observation is in order. Differently than the $4D$ case explored in
\cite{Gorbar:2002pw,Gorbar:2003yt}, here we could
establish the beta-function for the inverse Newton constant. The reason
for this difference is {\it not} the change of dimension, but rather the fact that
we used the heat kernel surface terms such as $G_R(-s\D)R$.
More details on how to obtain the surface terms can be found in \cite{Codello:2012kq,Avramidi:2007zz},
while non-trivial applications appear in \cite{El-Menoufi:2015cqw}.
In the next subsections we present a brief discussion of the properties of the main results
\eqref{gascal} and \eqref{betazero}.

\subsection{Recovering the Polyakov action in the conformal limit}
\label{s4.2}

We are now interested in the behavior of the effective action \eqref{gascal}
under the conformal limit $\xi\to 0$ and $m^2\to 0$. The term quadratic in the curvatures
of \eqref{gascal} can be understood as a generalization of the Polyakov action
\cite{Polyakov}, and therefore it is possible to see it as a toy model of
what one can expect in the more realistic $4D$ case for the non-conformal
scalar.

Let us start from the simplest case with $\xi=0$ and nonzero mass.
It is easy to see that the Polyakov action must be recovered in the limit
$z\to \infty$. This is the UV regime since $z$ diverges in the 
limit $m^2\to 0$ at fixed momenta $q^2$, or alternatively $q^2\to\infty$ at fixed $m^2$.
The function $C(z)$ was normalized so that it interpolates with the central charge
\beq
C(z) \to 1\,, \qquad {\rm for} \qquad
z\to \infty\,.
\eeq
As expected from local conformal invariance, in this limit the
quadratic term actually becomes Polyakov action
\beq\label{polyakov-limit}
\left.\Ga_{\rm s}[g]\right|_{R^2}
&\longrightarrow&
\,\frac{1}{96\pi} \int {\rm d}^2 x \sqrt{g}\,\, R \frac{1}{\D} R
\,=\, \Ga_{\rm P}[g]\,,
\eeq
in which we exclusively displayed the limit of the part of $\Ga_{\rm s}[g]$
that is quadratic in $R$.

In general, for a nonzero mass there are also infinitely many other
terms which display higher powers of $R$. The correspondence with
the Polyakov action  \eqref{polyakov-limit} requires that all these terms vanish in
the conformal limit, but the proof of this fact is beyond the scope
of the present article which deals only with the second order form
factors. Let us note, however, that the proof could constitute a relevant step,
particularly because of the interesting discussion of the role of
expansion in curvature tensor components in $4D$ that appeared in
\cite{DeserSchwimmer,Deser-2000}.

Even more interesting is to consider the theory with an arbitrary
value of $\xi$. Taking the UV limit in the general expression \eqref{gascal},
we can only expand about a small but nonzero mass $m^2$
and we arrive at a logarithmically
corrected Polyakov action
\beq\label{gaxi}
\left.\Ga_{\rm s}[g]\right|_{R^2}
&\longrightarrow&
\Ga_{\rm P}[g] \,-\,
\frac{\xi}{8\pi} \int {\rm d}^2 x \sqrt{g}\, R
\Big[
\frac{1}{\D}
\,-\, \xi \frac{\ln \left(-\D/m^2\right)}{\D}\Big] R\,.
\eeq
One can see that the massless limit of the
second term inside the integral is singular for $\xi \neq 0$.
Therefore the presence of a non-conformal value for $\xi$ forbids the massless limit.
At the same time in the conformal case $\xi = 0$ the result trivially interpolates
the Polyakov action as expected from \eqref{polyakov-limit}.
A more detailed discussion of the IR singularity for $\xi \neq 0$
can be found in Appendix \ref{appendix-non-analyticity}.

\subsection{Two extreme regimes for $\be_{G}^{\rm \,s}$}
\label{s4.3}

The beta fuction $\be_{G}^{\rm \,s}$ is a general expression which
is valid at all energy scales. Let us consider the UV regime with
$z\gg 1$, or equivalently $q^2 \gg m^2$; and the IR one with $z\ll1$,
or $q^2 \ll m^2$.  In these limits the vacuum beta function becomes
as in \eqref{UVIR},
\beq
\be^{\rm \,s}_{{G},\,{\rm UV}}
&=&
-\,\frac{1}{4\pi}\Big(\xi-\frac{1}{6}\Big)
- \frac{1}{4\pi}\,\frac{m^2}{q^2}
 \Big[ 1 - 2 \xi \ln \Big(\frac{q^2}{m^2}\Big) \Big]
 +\dots
\n{betaUV}
\\
\be^{\rm \,s}_{{G},\,{\rm IR}}
&=&
-\, \frac{1}{24\pi}\Big(\xi-\frac{1}{5}\Big) \frac{q^2}{m^2}
 + \dots
\n{betaIR}
\eeq
The first expression shows that at the leading order of the UV expansion
we recover the standard $\overline{\rm MS}$ result for $2D$
(see, e.g., \cite{birdav}),
\beq
\be^{\rm \,s}_{{G},\,{\overline{\rm MS}}}
&=&
-\,\frac{1}{4\pi}\Big(\xi - \frac16 \Big).
\n{betaMS}
\eeq
At the same time, in the IR we observe that for sufficiently small
$q^2$ the vacuum parameter stops running with a quadratic
decoupling term, as it should be according to the Appelquist and
Carazzone theorem \cite{AC}. It is remarkable that this result has
been achieved for the beta function of the $R$-term, that should be
regarded as a non-trivial result \cite{Gorbar:2002pw}.


\section{Dirac spinors}
\n{s5}

Let us denote the dimension of the Clifford algebra
$d_{\rm f} = \tr {\hat 1}$ and remember that in the general even
dimension $D$ we have $d_{\rm f} = 2^{D/2}$ (see, e.g., \cite{ParToms,MensAg}
for an introductions to fermions in curved space).
Furthermore, in $D=2$ the trace of the square of the commutator
of covariant derivatives is
\beq
 \tr \,\, \Omega_{\mu\nu}^2 &=& -\,\frac{d_{\rm f}}{8} \,\,R^2.
\eeq
Using these results and the general expression for the heat kernel
solution, we arrive at the fermion contribution in the form
\beq
\Ga_{\rm f}[g] 
&=&   
-\frac{1}{2} \Tr_{\rm f} \ln \Big(- \D + \frac{1}{4}R + m^2\Big)
\\
\nonumber
&=&
\,\frac{d_{\rm f}}{4\pi}\int {\rm d}^2 x \sqrt{g}\, \Big\{
\frac{1}{2\,\bar{\ep}}\Big(-m^2 - \frac{1}{12}\,R\Big)
- \frac{m^2}{2} + \frac12\,B_{\rm f}(z) R
+ \frac{1}{24}\, R\, \frac{C_{\rm f}(z)}{\D} \, R\Big\},
\nonumber
\eeq
in which
\beq
B_{\rm f}(z) \,=\, - \frac{1}{18}-\frac{2}{3} \frac{Y}{a^2}
\quad \mbox{and}\quad
C_{\rm f}(z) \,=\,
\frac{1}{2}\left(1-3Y\right)+\frac{6Y}{a^2}.
\eeq

Likewise the scalar case of the previous section, in the UV limit $m \to 0$ the form factor $C_{\rm f}$
interpolates the central charge of a single fermionic degree of freedom, namely $C_{\rm f}(z)\to 1/2$.
However, no running similar to that in Eq.~\eqref{gaxi} takes place in the case
of fermions (similarly with vectors, as we will show in the next
subsection), because in these cases there are no non-minimal
interactions with external metric such as $\xi$, and therefore there is no violation of {\it global}
scale invariance besides the mass.

The unique vacuum beta function is given by the expression
\beq
  \be_{G}^{\rm \,f}
&=&
\frac{d_{\rm f}}{8\pi}\,\Big\{\frac{1-3Y}{6} + \frac{2Y}{a^2}\Big\}.
 \eeq
In the UV ($q^2 \gg m^2$) and IR ($q^2 \ll m^2$) regimes we meet the
following limits:
\beq
\be^{\rm \,f}_{{G},\,{\rm UV}} &=&
\be^{\rm \,f}_{{G},\,{\rm \overline{MS}}}
\,+\, {\cal O}\Big(\frac{m^2}{q^2}\Big),
\quad
\mbox{with}
\quad
\be^{\rm \,f}_{{G},\,{\rm \overline{MS}}}
\,=\,
\frac{d_{\rm f}}{48\pi}\,,
\\
\be^{\rm \,f}_{{G},\,{\rm IR}}
&=&
\frac{d_{\rm f}}{480\pi} \frac{q^2}{m^2}
\,+\, {\cal O}\Big(\frac{q^2}{m^2}\Big)\,.
\eeq
As it was expected on the general grounds and in full analogy
with the $4D$ result \cite{Gorbar:2003yt}, the high-energy limit
shows a nice correspondence with the $\overline{\rm MS}$ scheme
beta function. At low energies we have once more a gravitational $2D$ version
of the decoupling theorem \cite{AC}. Once again, the remnant IR
running is related to the non-local terms which are of the
{\it second} order in curvature, while the divergence is a local
{\it first} order in curvature expression, in accordance with the
Weinberg theorem \cite{Weinberg-1960,Collins} and power counting.
This result is as remarkable as the one for the scalar field,
because in $4D$ it has not yet been achieved \cite{Gorbar:2002pw}.

\section{Massive and massless vector fields}
\n{s6}

For the massive Proca vector field one can use the general expression
\eqref{vect} and essentially repeat the computations described in the previous
section. We find
\beq
 \Ga_{\rm p}[g]
 &=&
 \frac{1}{2} \Tr_{\rm v} \ln \left( -\D + {\rm Ric}+m^2\right)
 - \frac{1}{2} \Tr_{\rm s} \ln \left(-\D +m^2\right)
\\
\nonumber
&=&
\frac{1}{4\pi}\int {\rm d}^2 x \sqrt{g}\, \Bigl\{
\frac{1}{2\,\bar{\ep}}\Big( m^2 + \frac{5}{6} R\Big)
+ \frac{m^2}{2}
+\frac{1}{2}\,B_{\rm p}(z) R
\,+\, \frac{1}{24}\, R\, \frac{C_{\rm p}(z)}{\D} R \Bigr\}\,,
\eeq
in which the form factors for the Proca field are defined
 \beq
 B_{\rm p}(z)
 &=&
 \frac{1}{18} + \frac{3}{2}\,Y + \frac{2Y}{3 a^2}
 \nonumber
 \\
 C_{\rm p}(z)
 &=&
 - \frac{1}{2} + 3 Y - \frac{3}{8}\, a^2 (Y-1) - \frac{6 Y}{a^2}\,.
\eeq
As one might expect from a standard counting of the degrees of freedom of a Proca field,
the function $C_{\rm p}(z)\to 1$ in the UV.

The beta function comes from the nonlocal term that is linear in the curvature scalar likewise \eqref{betazero}. It is given by
\beq
 \be_{G}^{\rm \,p}
&=&
\frac{1}{8\pi}\Big\{- \frac{1}{6} - Y + \frac38\, a^2 (Y-1) - \frac{2 Y}{a^2}\Big\}\,,
\n{beta-Proc}
\eeq
which is again a general expression valid at all energies scales. In the UV
($q^2 \gg m^2$) and IR ($q^2 \ll m^2$) it boils down to the simpler
results
\beq\label{Proca-UV}
\be^{\rm \,p}_{{G},\,{\rm UV}} &=&
\be^{\rm \,p}_{{G},\,{\overline{\rm MS}}}
\,+\, {\cal O}\Big(\frac{m^2}{q^2}\Big)\,,
\quad
\mbox{with}
\quad
\be^{\rm \,p}_{{G},\,{\overline{\rm MS}}}\,=\,
- \frac{5}{24\pi}\,,
\\
\label{Proca-IR}
\be^{\rm \,p}_{{G},\,{\rm IR}}
&=&
- \,\frac{1}{30\pi}\, \frac{q^2}{m^2}
\,+\, {\cal O}\Big(\frac{q^2}{m^2}\Big)\,.
\eeq
The last formula completes our determination of the $2D$ gravitational version of
the Appelquist and Carazzone theorem \cite{AC} including massive Proca fields.

In the UV limit \eqref{Proca-UV} one observes a difference with the well-known
result for the gauge massless vector
\beq
\be^{\rm \,g}_{{G},\,{\overline{\rm MS}}}\,=\,- \frac{1}{4\pi}\,.
\n{gauge}
\eeq
This difference is nothing else than the discontinuity in the massless
limit of the quantum contribution of the Proca field, which has been
discussed earlier in $4D$ \cite{Gorbar:2003yt,BuGui}
and we now observe in the $2D$ case.
The origin of this difference is in the fact that while a massive Proca field
requires only one ``compensating'' scalar degree of freedom, instead
a massless gauge field requires two, which are better known as Faddeev-Popov ghosts.

\section{Conclusions}
\label{s7}

The nonlocal form factors of the vacuum effective action have been derived
for several types of $2D$ massive fields in an approximation which includes
all the terms up to the second order in curvature. The
results have been obtained by means of the heat kernel approach.
An equivalent derivation based on Feynman diagrams
for the scalar field case is deferred to Appendix \ref{appendix-feynman}.

The form factors do not become logarithmic in the UV (or massless)
limit, because the divergences appear at the first order in the Ricci scalar.
However, one can recover the UV limit successfully, because the terms
of the second order in curvature reduce to the Polyakov action in the
case of $m \to 0$ for fermions and vectors, while for the scalar field
such a limit is achieved only for the minimal version, with $\xi=0$,
which is known to be classically conformally invariant in $2D$.

In the massless limit of the the nonminimal scalar we find a
modified version of the Polyakov action, which includes a qualitatively
new logarithmic form factor. The new logarithm can speculatively
be interpreted as a kind of IR renormalization group running that does not have a direct
relation to the UV divergences.

In the $\overline{\rm MS}$ renormalization
scheme $\be$-functions related to the divergences that appear in the
linear part in the curvature are reproduced as the UV limit of more
general expressions corresponding to the momentum subtraction
scheme. At the opposite end of the energy scale, in the IR, we
meet a $2D$ version of the gravitational decoupling theorem
very similar to the $4D$ counterpart \cite{Gorbar:2002pw,Gorbar:2003yt}.
The main difference between the results in $4D$ and $2D$ is that
in the latter case the UV limit reproduces the anomaly-induced action
and IR limit shows decoupling in the beta functions {\it without} the
usual correlation between divergences and logarithmic form factors.

We believe that our result are instructive for better understanding
how one can explore the running of Newton and cosmological
constants, including the cases in which the form factors are
formally irrelevant \cite{Gorbar:2002pw}.
Let us remind the Reader that the question on whether there is a remnant IR running of these
parameters or not has potentially interesting cosmological applications
\cite{PoImpo,DCCR} and can be relevant in astrophysics too
(see, e.g., \cite{RotCurves,Rodrigues:2012qm}). For these reasons the
present work can be seen as a small step forward in understanding
how the information concerning the part of the vacuum effective action that is linear in the curvature
can be used to interpolate between UV and IR physics through a renormalization group running
based on a physical scale.

\appendix

\section{Non-local heat kernel expansion}\label{appendix-heat-kernel}

In this appendix we provide more details on the form of the non-local expansion of the heat kernel
which was originally developed in \cite{bavi87,bavi90}.
We follow however the notation of \cite{Codello:2012kq}. Consider an operator of Laplace type
\beq
{\cal O} \,=\, - \, \D + E\,,
\eeq
acting on a general vector bundle which equipped with a connection
and which lives over a Riemaniann manifold with metric $g_{\mu\nu}$.
The operator is the sum of the Laplacian $\D=\nabla^2=g^{\mu\nu}\nabla_\mu\nabla_\nu$
and an endomorphism over the bundle. In general the connection has curvature
$\Omega_{\mu\nu}=\left[\nabla_\mu,\nabla_\nu\right]$ and might even contain a Levi-Civita contribution
if part of the vector bundle is obtained as a tensor product of tangent or co-tangent bundles. 

The heat kernel ${\cal H}(s;x,x')$ is defined as the solution of
the following Cauchy problem with respect to the proper time $s$:
\beq
&&
\left(\pa_s + {\cal O}_x\right) \, {\cal H}(s;\,x,x') \,=\, 0
\nn
\\
&&
{\cal H}(s;\,x,x') \,=\, \de(x,x')
\n{Cauchy}
\eeq
with $\de(x,x')$ the Dirac delta over the manifold \cite{Avramidibook}.

The trace of the coincidence limit of the heat kernel ${\cal H}(s;\,x,x)$ admits an expansion in terms of Riemaniann and connection's curvatures and of the endomorphism.
The expansion can be computed unambiguously over asymptotically flat manifolds \cite{bavi90,Avramidi:2007zz,Codello:2012kq}.
Up to second order in the curvatures it has the form
\beq
 \tr ~ {\cal H}(s,x,x)
 &=&
 \frac{1}{(4\pi s)^{D/2}}\int{\rm d}^D x \sqrt{g}\,
 \tr \Big\{ \mathbf{1}\,
  + s \big[G_E(-s\D) E   +  G_R(-s\D) R\big]
  \nonumber
  \\
  && + s^2 \Big[ R F_R(-s\D)R
  + R^{\mu\nu} F_{Ric}(-s\D)R_{\mu\nu}
  + E F_E(-s\D)E
  \\
 && + E F_{RE}(-s\D)R
  + \Omega^{\mu\nu} F_\Omega(-s\D) \Omega_{\mu\nu}\Big]
 \Big\}
  + {\cal O}\left({\cal R}^3\right)\,,
\nonumber
\eeq
in which ${\cal O}\left({\cal R}^3\right)$ indicates a nonlocal, but consistent, curvature
expansion to the third order as well explained in \cite{bavi90}.
The above formula is given without specific boundary conditions at infinity for the first order in the
curvatures, while integration by parts is used for the second order
(see \cite{Codello:2012kq,El-Menoufi:2015cqw} for more details on this derivation).

The nonlocal functions of $\Delta_g$ appearing in the expansion
are known as form factors of the heat kernel. They are obtained as
\beq
 G_E(x) &=& -f(x)\,,
 \n{ff1}
 \\
 G_R(x) &=& \frac{f(x)}{4}+\frac{f(x)-1}{2x} \,,
 \n{ff2}
 \eeq
for the terms that are linear in the curvatures, and
\beq
 F_{Ric}(x) &=& \frac{1}{6x}+\frac{f(x)-1}{x^2}\,,
 \n{ff3}
 \\
 F_R(x)
 &=&
 -\frac{7}{48x}+\frac{f(x)}{32}+\frac{f(x)}{8x}-\frac{f(x)-1}{8x^2} \,,
 \n{ff4}
\\
F_{RE}(x)
&=&
-\frac{f(x)}{4}-\frac{f(x)-1}{2x} \,,
\n{ff5}
\\
F_E(x) &=&\frac{f(x)}{2} \,,
\n{ff6}
\\
F_\Omega(x) &=& -\frac{f(x)-1}{2x} \,,
\eeq
for the terms which are quadratic in the curvatures.
All form factors are expressed via the basic form factor
\beq\label{ffunction}
 f(x) &=& \int_0^1 \!{d}\alpha \, {\rm e}^{-\alpha(1-\alpha)x}\,.
\eeq
These form factors admit well-defined expansions both for large and
small values of the parameter $s$ (see the discussion of Appendix \ref{appendix-non-analyticity}),
and are used in the main text to obtain the related nonlocal structures
which appear in the vacuum effective action of matter fields.

\section{Curvature renormalization from Feynman diagrams}\label{appendix-feynman}

The standard perturbative approach to the renormalization of surface
terms such as power series composed by $(-\D)^n R$ is particularly
difficult if Feynman diagrams and an expansion of the metric
around flat space $g_{\mu\nu}=\delta_{\mu\nu}+h_{\mu\nu}$ are
adopted. Let us comment on the nature of this difficulty and
how we solve it in the present
work. The interested reader can consult \cite{Codello:2012kq} for
more details.

Momentum conservation generically constrains external lines in the fluctuation
$h_{\mu\nu}$ to have zero incoming momentum in one-point functions.
This happens, in particular, for diagrams which would
otherwise renormalize the scalar curvature itself.
In fact, the first vertex in
$\,h_{\mu\nu}\,$ is
\beq
\frac{1}{\sqrt{g}}\frac{\de}{\de h_{\mu\nu}}
\int d^2x \sqrt{g} R\bigg|_{\de_{\mu\nu}}
\,=\,
\int \frac{d^2p}{(2\pi)^2}\,e^{ipx}
\,\big(p^2 g^{\mu\nu}-p^\mu p^\nu\big).
\eeq
The first tensor structure in this expression receives contributions
from the one-point function of the volume term too (cosmological
constant), but the second one is only sensitive to operators such as
$(-\D)^n R$ (see Appendix of \cite{Codello:2012kq}).
It is clear that one cannot see any renormalization of this vertex if
the following diagram is considered
\begin{equation}
\begin{split}
 \myincludegraphics{33}{0.40}{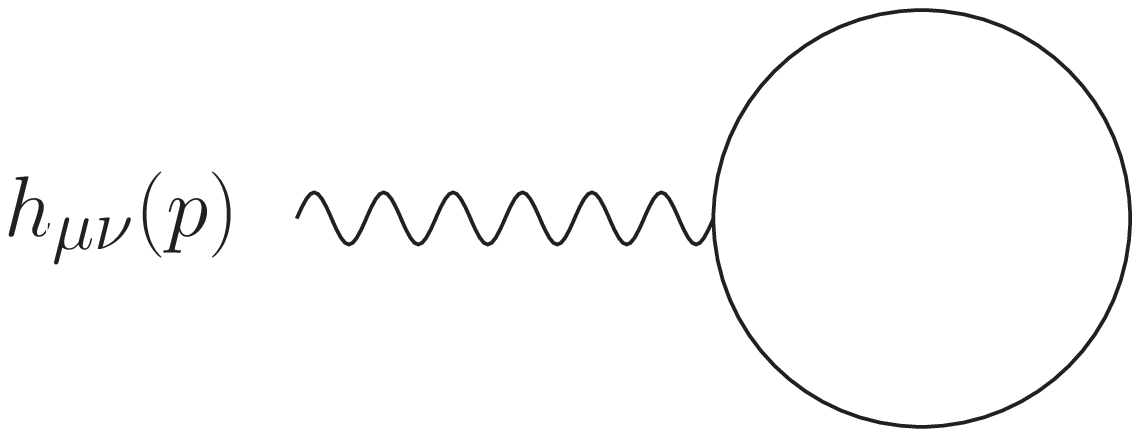} 
\end{split}
\n{one-point}
\end{equation}
and the momentum $p_\mu$ is constrained to be zero by the
conservation law.

In order to be able to retain a nonzero incoming momentum $p_\mu$
we insert a new ``identity'' vertex $\mathbf{1}$ in the theory
which is trivially defined by momentum conservation.
Then, instead of \eqref{one-point} we can consider the following diagram:
\begin{equation}\label{2PFhmunu-id}
\begin{split}
 \myincludegraphics{33}{0.40}{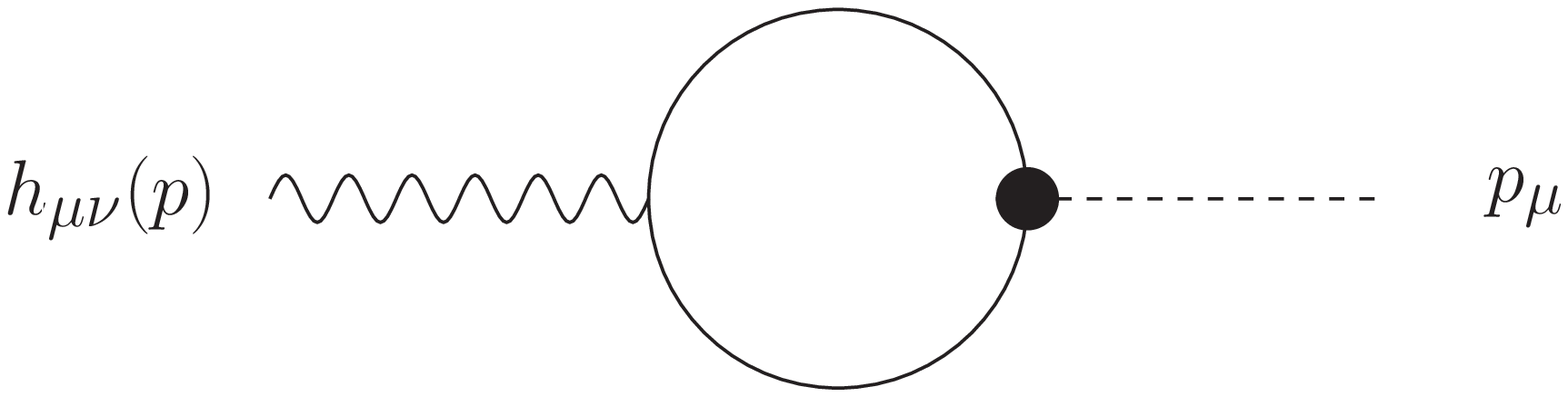}
\end{split}
\end{equation}
In coordinate space this diagram contributes to
\begin{equation}
\begin{split}
 \int {\rm d}^d x \sqrt{g} \,\mathbf{1} \, g(-\D) R,
\end{split}
\end{equation}
in which $g(-\D)$ is a function to be uniquely
determined by loop integration of the diagram itself.
E.g., taking the limit $\mathbf{1}\to 1$ in the scalar case one
can use the new diagram to renormalize the nonlocal contribution to
the Einstein-Hilbert term.


The integral can be easily computed introducing a Feynman parameter $0\le\alpha\le 1$ and exponentiating
the combined propagators with a proper time parameter $s$.
We display only the minimal $\xi=0$ case for the sake of simplicity.
The final result is an integral over the parameters $s$ and $\alpha$
\beq
{\rm Eq.}~\eqref{2PFhmunu-id}
\,\,\propto \, \,
 I(p^2) \, p_\mu p_\nu +\dots\,\,,
\eeq
in which we determine
\beq
 I(p^2)\,=\,
 -\frac{1}{(4\pi)^{d/2}} \int {\rm d}s\, s^{1-d/2}\, \, {\rm e}^{-m^2s}
 \, \Big\{ \frac{1}{2sp^2}
 - \frac{2+sp^2}{4sp^2}f(sp^2)\Big\}\,.
\n{int}
\eeq
In this expression the dots stand for further contributions proportional to
the background metric (which are needed for the renormalization of the
cosmological constant) and $f(x)$ is the basic form factor \eqref{ffunction} of the
heat kernel. The function
\beq
 I(p^2)
 =
  \int{\rm d}s ~ \frac{1}{(4\pi s)^{d/2}} s ~ G_R(sp^2)
\eeq
gives the nonlocal contribution to the Einstein-Hilbert term. One can see that the integrand
is exactly the form factor associated with $R$ in the nonlocal
expansion \eqref{ff2} (see also \cite{Avramidi:2007zz}).
%
From this point onward, the computation of the vacuum effective action can
straightforwardly follow the main text because it coincides with what is obtained from the heat kernel expansion.

\section{On the non-analyticity of expression \eqref{gaxi}} \label{appendix-non-analyticity}

In this appendix we elaborate in more detail the reasons
why the integrand of \eqref{gaxi} is singular in the massless limit.
We follow the steps of Refs.~\cite{bavi87,bavi90}
in which it is shown that the effective two dimensional
action is analytical only in the case of a
conformal scalar field.

The analyticity of the effective action is related to the convergence of
the integral \eqref{gamma} in the upper limit. In this limit the behavior
of  \eqref{hk} is determined by the expansion of the form factors
to large values of $s$. From the expressions \eqref{ff1} - \eqref{ff6}
one can obtain the following asymptotic behaviors of the form
factors:
\beq
 G_E(x) &=& -\frac{2}{x}+\mathcal{O}\Big(\frac{1}{x^2}\Big)\,,
 \\
 G_R(x) &=& 0+\mathcal{O}\Big(\frac{1}{x^2}\Big) \,,
 \eeq
and
\beq
 F_{Ric}(x) &=& \frac{1}{6x}+\mathcal{O}\Big(\frac{1}{x^2}\Big)\,,
 \\
 F_R(x)
 &=&
 -\frac{1}{12x}+\mathcal{O}\Big(\frac{1}{x^2}\Big) \,,
\\
F_{RE}(x)
&=&
0+\mathcal{O}\Big(\frac{1}{x^2}\Big) \,,
\\
F_E(x) &=& \frac{1}{x}+\mathcal{O}\Big(\frac{1}{x^2}\Big) \,.
\eeq
Combining everything together,
the behavior of $\mathcal{H}(s)$ for $s\rightarrow\infty$ is
\beq
\mathcal{H}(s)
&=&
s^{-\frac{D}{2}}(\mathcal{R})^{0}+s^{-\frac{D}{2}}(\mathcal{R})^{1}
+s^{-\frac{D}{2}+1}(\mathcal{R})^{2}+\mathcal{O}(\mathcal{R})^{3},
\eeq
in which $\mathcal{R}$ symbolically represents all kinds of curvature and indices
$0$, $1$, $2$ determine the order of expansion.

Comparing the previous expression to $\mathcal{H}(s)$ with
Eq.~(2.16) in Ref.~\cite{bavi90}, we note that the presence of the
form factor in the linear part in curvature changes the behavior of
$\mathcal{H}(s)$ at large $s$. In the computation of $\Gamma[g]$
the coefficient of the linear term in the curvature is determined by an
integral of the type
\begin{equation}
\int^{\infty}\frac{{\rm d}s}{s}\frac{1}{(4\pi s)^{D/2}}\Big\{1+\mathcal{O}\Big(\frac{1}{s}\Big)\Big\}\,,
\end{equation}
while for the quadratic term we have
\begin{equation}
\int^{\infty}\frac{{\rm d}s}{s}\frac{s}{(4\pi s)^{D/2}}\Big\{1+\mathcal{O}\Big(\frac{1}{s}\Big)\Big\}\,,
\label{intdiv}
\end{equation}
when we consider $m=0$.
For $D = 2$ the effective action is analytic until the first order
in curvature. On the other hand, in the second order in curvature
the analyticity of $\Gamma[g]$ is broken by the presence of the
term $\,1/s\,$ in  the integrand. It is this kind of singularity that
appears in the second term of the integral \eqref{gaxi}.

From this perspective the Polyakov action is a very special case,
since it is analytic at second order in the curvature. The reason is
that the form factors combine such that the terms behaving like
$\,1/s\,$ do cancel and the first term of the expansion for a large
$s$ becomes $\,1/s^2$, making $\Ga[g]$ finite in the IR. From
$\mathcal{H}(s)$ and the expansion of the form factors for large
$s$, one can see that the cancellation of the $\,1/s\,$ terms occurs
only when $\xi=0$. Only in this case the $m=0$ effective action is
analytic at $\,D=2$. On the contrary, the presence of the term with
$\,1/s\,$ in the case $\xi\neq 0$ leads to  a function $\Ga(1-D/2) $,
which has a pole at $D=2$. This pole corresponds to the IR
divergence related to the singularity of the effective action in $2D$
\cite{bavi90,Avramidibook}.


\end{document}